\documentclass[structabstract]{aa}

\usepackage{graphicx}      
\usepackage{longtable}      
\usepackage{lscape}      
\usepackage{txfonts}
\usepackage{natbib}
\usepackage{url}
\usepackage{numprint}		
\usepackage[squaren,cdot]{SIunits}	
\usepackage{mathrsfs}		
\usepackage{xspace} 			

\newcommand{\he}{\object{HE0435--1223}\xspace} 

\bibpunct{(}{)}{;}{a}{}{,} 

\begin{document}
\title{I. Flux and color variations \\ of the quadruply imaged quasar
  HE 0435--1223 \thanks{Based on data collected by MiNDSTEp with the
    Danish 1.54m telescope at the ESO La Silla Observatory.  Tables 5,
    6, and 7 are only available in electronic form at the CDS via
    anonymous ftp to cdsarc.u-strasbg.fr (130.79.128.5) or via
    http://cdsweb.u-strasbg.fr/cgi-bin/qcat?J/A+A/} }
\author{
  D.~Ricci\inst{1}\fnmsep\thanks{Boursier FRIA} \and
  J.~Poels\inst{1}\and
  A.~Elyiv\inst{1}\fnmsep\inst{2}\and
  F.~Finet\inst{1}\and
  P.~G.~Sprimont\inst{1}\and
  T.~Anguita\inst{3}\fnmsep\inst{4}\and 
  V.~Bozza\inst{5}\fnmsep\inst{6}\and 
  P.~Browne\inst{7}\and 
  M.~Burgdorf\inst{8}\fnmsep\inst{24}\and
  S.~Calchi Novati\inst{5}\fnmsep\inst{9}\and 
  M.~Dominik\inst{7}\fnmsep\thanks{Royal Society University Research Fellow}\and 
  S.~Dreizler\inst{10}\and 
  M.~Glitrup\inst{11}\and 
  F.~Grundahl\inst{11}\and 
  K.~Harps\o e\inst{12}\fnmsep\inst{25}\and 
  F.~Hessman\inst{10}\and 
  T.~C.~Hinse\inst{12}\fnmsep\inst{13}\and 
  A.~Hornstrup\inst{14}\and 
  M.~Hundertmark\inst{10}\and 
  U.~G.~J\o rgensen\inst{12}\fnmsep\inst{25}\and 
  C.~Liebig\inst{7}\fnmsep\inst{15}\and 
  G.~Maier\inst{15}\and 
  L.~Mancini\inst{5}\fnmsep\inst{9}\fnmsep\inst{16}\and 
  G.~Masi\inst{17}\and 
  M.~Mathiasen\inst{12}\and 
  S.~Rahvar\inst{18}\and 
  G.~Scarpetta\inst{5}\fnmsep\inst{6}\and 
  J.~Skottfelt\inst{12}\and 
  C.~Snodgrass\inst{19}\fnmsep\inst{20}\and 
  J.~Southworth \inst{21}\and 
  J.~Teuber\inst{12}\and 
  C.~C.~Th\"one\inst{22}\fnmsep\inst{23}\and 
  J.~Wambsgan\ss\inst{15}\and 
  F.~Zimmer\inst{15}\and 
  M.~Zub\inst{15} 
  \and
  J.~Surdej\inst{1}\fnmsep\thanks{also Directeur de Recherche honoraire du FRS-FNRS}
}
\institute{
  D\'epartement d'Astrophysique, G\'eophysique et Oc\'eanographie, B\^at.~B5C, Sart Tilman,
  Universit\'e de  Li\`ege, 
  B-4000 Li\`ege 1, Belgique;
  \email{ricci@astro.ulg.ac.be}
  \and 
  Main Astronomical Observatory, Academy of Sciences of Ukraine, Zabolotnoho 27, 03680 Kyiv, Ukraine
  \and 
  Centro de Astro-Ingenier\'ia, Departamento de Astronom\'ia y Astrof\'isica, Pontificia Universidad Cat\'olica de Chile, Casilla 306, Santiago, Chile.
  \and 
  Max-Planck-Institut f\"ur Astronomie, K\"onigstuhl 17, 69117 Heidelberg, Germany
  \and 
  Dipartimento di Fisica ``E.R. Caianiello'', Universit\`a degli Studi di Salerno, Via Ponte Don Melillo, 84085 Fisciano (SA), Italy
  \and 
  Instituto Nazionale di Fisica Nucleare, Sezione di Napoli, Italy
  \and 
  SUPA, University of St~Andrews, School of Physics \& Astronomy, North Haugh, St~Andrews, KY16 9SS, UK
  \and 
  Deutsches SOFIA Institut, Universitaet Stuttgart, Pfaffenwaldring 31, 70569 Stuttgart, Germany
  \and 
  Istituto Internazionale per gli Alti Studi Scientifici (IIASS), Vietri Sul Mare (SA), Italy
  \and 
  Institut f\"ur Astrophysik, Georg-August-Universit\"at G\"ottingen, Friedrich-Hund-Platz 1, 37077 G\"ottingen, Germany
  \and 
  Department of Physics \& Astronomy, Aarhus University, Ny Munkegade, 8000 Aarhus C, Denmark
  \and 
  Niels Bohr Institute, University of Copenhagen, Juliane Maries vej 30, 2100 Copenhagen \O, Denmark
  \and 
  KASI - Korea Astronomy and Space Science Institute, 61-1 Hwaam-dong, Yuseong-gu, Daejeon 305-348, Republic of Korea
  \and 
  National Space Institute, Technical University of Denmark, 2800 Lyngby, Denmark
  \and 
  Astronomisches Rechen-Institut, Zentrum f\"ur Astronomie, Universit\"at Heidelberg, M\"onchhofstra\ss e 12-14, 69120 Heidelberg, Germany
  \and 
  Dipartimento di Ingegneria, Universit\`a del Sannio, Corso Garibaldi 107, 82100 Benevento, Italy
  \and 
  Bellatrix Astronomical Observatory, Center for Backyard Astrophysics, Ceccano (FR), Italy
  \and 
  Physics Department, Sharif University of Technology, Tehran, Iran
  \and 
  European Southern Observatory, Casilla 19001, Santiago 19, Chile
  \and 
  Max Planck Institute for Solar System Research, Max-Planck-Str. 2, 37191 Katlenburg-Lindau, Germany
  \and 
  Astrophysics Group, Keele University, Newcastle-under Lyme, ST5 5BG, UK
  \and 
  Dark Cosmology Centre, Niels Bohr Institute, University of Copenhagen, Juliane Maries Vej 30, Copenhagen Ø, 2100 Denmark  
  \and 
  INAF, Osservatorio Astronomico di Brera, 23807 Merate, Italy
  \and 
  SOFIA Science Center, NASA Ames Research Center, Mail Stop N211-3, Moffett Field CA 94035, USA
  \and 
  Centre for Star and Planet Formation, Geological Museum, \O ster Voldgade 5, 1350 Copenhagen, Denmark.
}


\abstract
    {}
    { We present $VRi$ photometric observations of the quadruply
      imaged quasar \he, carried out with the Danish $1.54\meter$ telescope
      at the La Silla Observatory. Our aim was to monitor and study
      the magnitudes and colors of each lensed component as a function
      of time. }
    { We monitored the object during two seasons (2008 and 2009) in
      the $VRi$ spectral bands, and reduced the data with two
      independent techniques: difference imaging and PSF (Point Spread
      Function) fitting. }
    { Between these two seasons, our results show an evident decrease in
      flux by $\approx$ 0.2--0.4 magnitudes of the four lensed components in
      the three filters. We also found a significant increase
      ($\approx$ 0.05--0.015) in their $V-R$ and $R-i$ color
      indices. }
    { These flux and color variations are very likely caused by
      intrinsic variations of the quasar between the observed
      epochs. Microlensing effects probably also affect the brightest
      ``A'' lensed component.}
    
    \maketitle
    
\section{Introduction}

\begin{figure}[t]
  \centering \includegraphics[width=8.7cm]{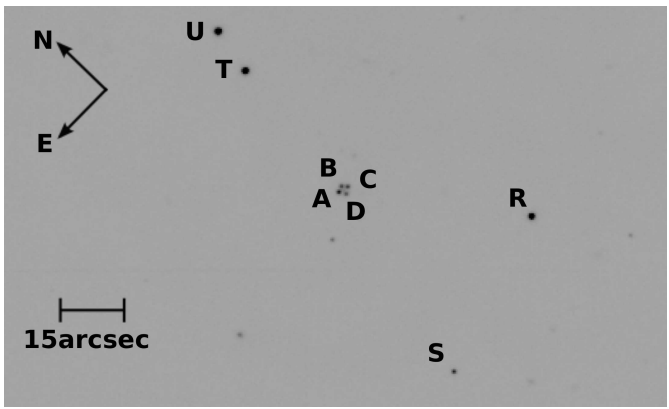}
  \caption{Zoom of a DFOSC $i$ filter image showing the four
    components of the lensed quasar and four nearby stars. The
    components are labeled following the notation of
    \cite{wisotzki02}: ``A'' for the brighter component, ``B'', ``C''
    and``D'' clockwise. The stars ``R'', ``S'', ``T'' and ``U'' were
    used to search for a suitable reference star. The ``R'' star was
    finally chosen. The contrast of the displayed image, on a negative
    scale, was selected to improve the visibility of the lensed
    components. The image is a median of the three CCD frames
    collected on 2008 August 8.}
  \label{fig:mediumfield}
\end{figure}

In the framework of the MiNDSTEp (Microlensing Network for the
Detection of Small Terrestrial Exoplanets) campaign \citep{dominik10},
which has as a main target the systematic observation of bulge
microlenses, we developed a parallel project concerning photometric
multi-band observations of several lensed quasars\footnote{\he,
  UM673/\object{Q0142--100}, \object{Q2237+0305},
  \object{WFI2033---4723} and \object{HE0047--1756}}. In the present
paper we focus on \he (see~Fig.~\ref{fig:mediumfield}), a QSO
discovered by \cite{wisotzki00} in the course of the Hamburg/ESO
digital objective prism survey, and confirmed to be a quadruply imaged
quasar by \cite{wisotzki02}. The lensing galaxy was initially
identified as an elliptical with a scale length of $\approx12\kilo\rm
pc$ at a redshift in the range $z=0.3$--$0.4$. The time delays between
the four images (labeled ``A'', ``B'', ``C'', ``D'', starting from the
brighter one and proceeding clockwise) of the quasar were estimated
around 10 days, and the quasar itself showed some signs of intrinsic
variability \citep{wisotzki02}.

\noindent More recently, the value of the redshift for the lensing galaxy
was estimated as $z=0.44\pm 0.20$, and the quasar redshift was
confirmed to be $z=1.6895\pm 0.0005$, with a $\Delta z$ between the
components of $\approx 0.0015$ rms \citep{wisotzki03}. These
spectrophotomeric observations showed some possible signature of
microlensing effects in the continuum and in the spectral emission
lines for the ``D'' component.

\begin{figure}[t]
  \centering
  \includegraphics[width=8.7cm]{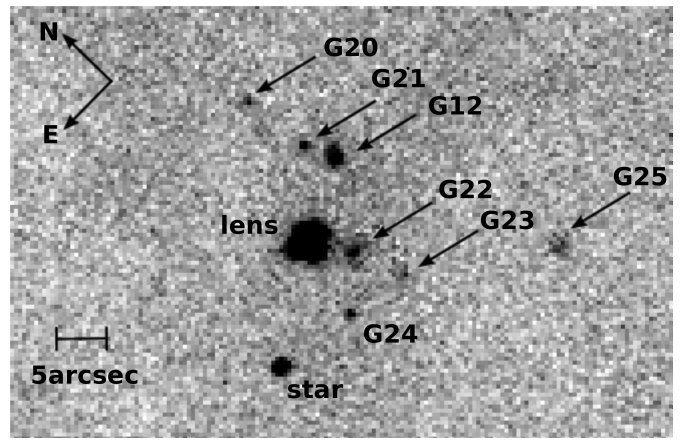}
  \caption{Zoom of a DFOSC $i$ filter image showing the galaxy
    environment near \he. The objects are labeled following the
    notation of \cite{morgan05}. The contrast, on a negative scale,
    was selected to improve the visibility of the galaxies. The image is
    a median of the three CCD frames collected on 2008 August 8. }
  \label{fig:bigfield}
\end{figure}

\noindent \cite{morgan05} provided milliarcsecond astrometry, revised
the value of the lens redshift at $z=0.4546\pm 0.0002$ with the
Low-Dispersion Survey Spectrograph 2 (LDSS2) on the Clay $6.5\meter$
telescope, and studied the galaxy environment of the lens, because it is
located in a dense galaxy field. The results do not show any evidence
of a cluster for the considered galaxies. However, the nearest
galaxies (G20, G21, G22, G23, and G24 in Fig.~\ref{fig:bigfield}) whose
redshifts were not measured, left this scenario open. Nevertheless, the
results of a deep investigation concerning the direction of an
external shear in the gravitational field of the lens do not show an
evident correlation with the position of the near galaxies. As a
remaining explanation, \cite{morgan05} suggested the presence of
substructures in the lensing galaxy.

\noindent The first systematic monitoring, which was performed in the
$R$ filter and covered the years between 2003 and 2005, was carried
out by \cite{kochanek06}: that paper provided astrometric measurements
compatible with the previous works,  measured the time delays between
the images ($\Delta t_{AD}= -14.37$, $\Delta t_{AB}= -8.00$, and
$\Delta t_{AC}= -2.10$ days, with errors respectively of $6\%$, $10\%$
and $35\%$), and finally confirmed the lensing galaxy as an elliptical
with a rising rotation curve.

\noindent Furthermore, \cite{mediavilla09} observed \he in the
framework of a monitoring of 29 lensed quasars, and attributed
eventual microlensing events to the normal stellar populations, while
\cite{blackburne10a} focused on the quasar itself, applying a model
with a time-variable accretion disk to the object. \cite{mosquera10}
found clear evidence of chromatic microlensing in the ``A'' component,
and provided an estimate of the disk size in the $R$ band in agreement
with the simple thin-disk model. \cite{blackburne10b} used the
chromatic microlensing to model the accretion disk, and
\cite{courbin10} recalculated the time delays with N-body realizations
of the lensing galaxy, which he thought to belong to the ``B
component'' ($\Delta t_{BA}= 8.4$, $\Delta t_{BC} =7.8$ and $\Delta
t_{BD}= −6.5$ days with errors of $25\%$, $10\%$, and
$11\%$ respectively).
Considering multi-color observations of other lensed quasars, a
single-epoch multi-band photometry was used on \object{MG0414+0534} to
constrain the accretion disk model and the size of the emission region
in the continuum \citep{bate08a,bate08b,floyd08}.
\noindent A multi-epoch multi-band photometry, carried out during
several years, was used for the quasar \object{Q2237+0305} by
\cite{kopletova06}, who observed the object during five years
(1995--2000) in the $VRI$ bands. \cite{anguita08} combined these
data with OGLE observations. \cite{mosquera09} monitored the
object in eight filters and found evidence for microlensing in the
continuum, but not in the emission lines.
\noindent Furthermore, \object{Q2237+0305} was the object of deep
studies focused on the lens galaxy \citep{poindexter10a}, and on the
inclination of the accretion disk \citep{poindexter10b}.
%
Another example of multi-epoch multi-band observations is given by
UM673/\object{Q0142--100}, observed in the Gunn $i$ and Cousins $V$
filters between 1998 and 1999 \citep{nakos05} and in the $VRI$ bands
between 2003 and 2005 \citep{kopletova09}.
\noindent Unlike for these objects, no \emph{systematic} multi-band
photometry has ever been carried out for \he.

Here, we present two periods of multi-band photometric observations of
\he, performed in the $VRi$ spectral bands with the Danish
$1.54\meter$ telescope at the La Silla Observatory. 

\noindent In Sect.~\ref{sec:obs} we explain how the observations were
carried out; in Sect.~\ref{sec:red} we focus on the data reduction,
and we describe the two independent techniques: difference imaging and
PSF (Point Spread Function) fitting, that were used to construct the
light curves. In Sect.~\ref{sec:res} we present the results. Finally,
in Sect.~\ref{sec:conc} we summarize the conclusions.

\section{Observations and pre-processing}
\label{sec:obs}
We observed \he during two seasons (2008 and 2009) with the Danish
$1.54\meter$ telescope at the La Silla Observatory. We used the DFOSC
instrument (Danish Faint Object Spectrograph and Camera) for imaging
and photometry, with a $2147\times 2101$ CCD device, covering a
$13.7\arcmin\times13.7\arcmin$ field of view with a resolution of
$0.39\arcsec/\rm pixel$. The gain of the device is $0.74$
electron/ADU in high mode, while the read out noise in this
mode is $3.1$ electrons \citep{sorensen00}.

\begin{figure}[t]
  \centering
  \includegraphics[width=8.7cm]{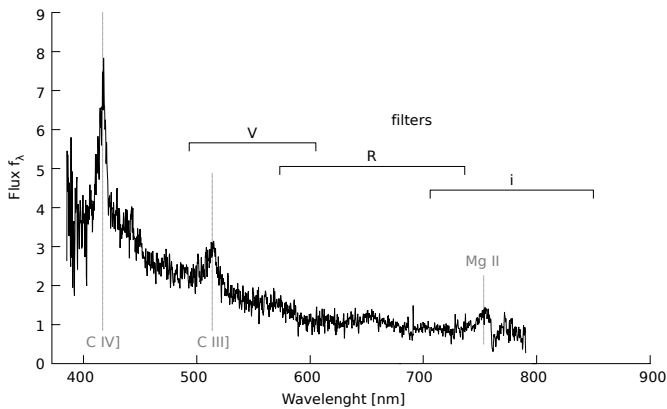}
  \caption{Position and bandwidth of the $VRi$ DFOSC filters
    superposed on the unresolved spectrum of \he. The figure is a
    composition made from the spectrum shown in the paper of
    \cite{wisotzki02}.}
  \label{fig:lambda}
\end{figure}

\noindent The data were collected in three different filters: Gunn $i$,
Bessel $R$ and Bessel $V$ (see Table~\ref{tab:filters} and
Fig.~\ref{fig:lambda}). We worked with a very homogeneous dataset
consisting of $180\second$ exposures.

\noindent For almost every night of observation, we also collected
bias images and dome flat-fields, which were already treated
\textit{in loco} using an automatic IDL procedure, part of the
MiNDSTEp pipeline for the observation of bulge microlenses. We then
obtained master flat-fields for the different filters and master
biases. When these images were not present for the desired date, we
coupled the most recent set of master bias and master flat-fields to
our science dataset, in the phase of pre-processing.

\begin{table}[b]
  \caption{Parameters of the DFOSC filters used for the
    photometry
    \setcounter{footnote}{2}\protect\footnotemark[\value{footnote}].}
  \label{tab:filters}
  \centering
  \begin{tabular}{lcccc}
    \hline\hline 
    Filter              &  ESO  &  Size           & $\lambda$      & $\Delta\lambda$ \\
    &  \#   & [\milli\meter]  & [\nano\meter]  & [\nano\meter] \\
    \hline	
    Bessel $V$   &  451  &   C60.0         &   544.80       & 116.31  \\
    Bessel $R$   &  452  &   C60.0         &   648.87       & 164.70 \\
    Gunn $i$     &  425  &   C60.0         &   797.79       & 142.88\\
    \hline
  \end{tabular}
\end{table}
\footnotetext{Information available on the Internet at
  \url{http://www.eso.org/lasilla/telescopes/d1p5/misc/dfosc_filters.html}}

\noindent We collected a total of 391 images during the 2008 season,
and 160 images in 2009.

\noindent The images were pre-processed (de-biased and flat-fielded),
and we used particularly the dome flats. To erase the possible
residual halos caused by the inhomogeneous illumination, the sky
background was subtracted fitting a $4^{th}$ degree surface after
masking the stars and cosmic rays, and the images were recentered
with an accuracy of 1px.

\noindent These steps were performed with a C++ pipeline developed
by our team.

\noindent We analyzed each image to sort out and then
disregard the problematic images in terms of bad tracking,
particularly bad seeing (the components were completely unresolved)
or bad focusing.

\noindent We then obtained 216 images during the 2008 season:
70 in the $i$ filter, covering 26 nights, 
83 in the $R$ filter, covering 32 nights, and 
63 in the $V$ filter, covering 25 nights, 
distributed between 2008 July 27 and 2008 October 4.  \\
Concerning the 2009 season, we obtained 116 images:
46 in the $i$ filter, covering 17 nights, 
37 in the $R$ filter, covering 14 nights, and 
33 in the $V$ filter, covering 12 nights, 
distributed between 2009 August 20 and 2009 September 19.

\section{Data reduction}
\label{sec:red}
As a first step, we chose four stars near \he, labeled ``R'', ``S'',
``T'' and ``U'' in Fig.~\ref{fig:mediumfield}, to search for a stable
reference star. We examined the ratios between the fluxes of these
stars in the $V$ band as a function of time, to possibly detect some
photometric variations between the two seasons.
\begin{figure}[t]
  \centering
  \includegraphics[width=8.7cm]{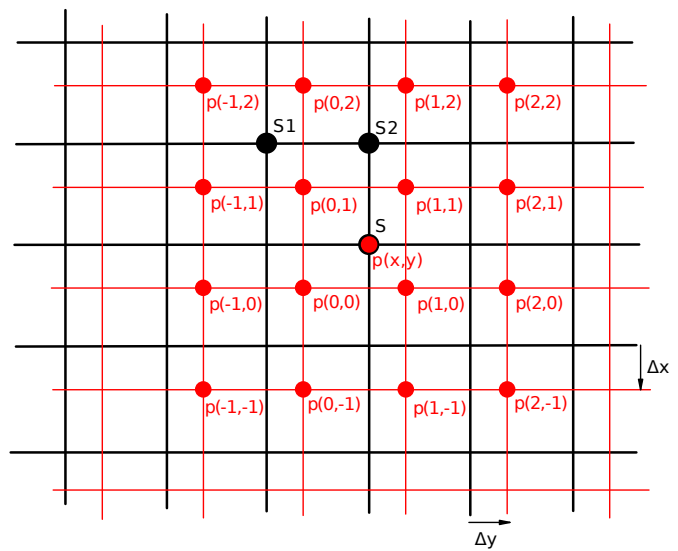}
  \caption{Scheme for the superposed images of the first (little dots, lighter grid) and
    second (big dots, bolder grid) stars. The big dots correspond to the knots with
    known fluxes. The arrows $\Delta x$ and $\Delta y$ indicate the
    shift between the two images.}
  \label{fig:grid}
\end{figure}

\begin{figure*}[t]
  \centering
  \includegraphics[width=17.4cm]{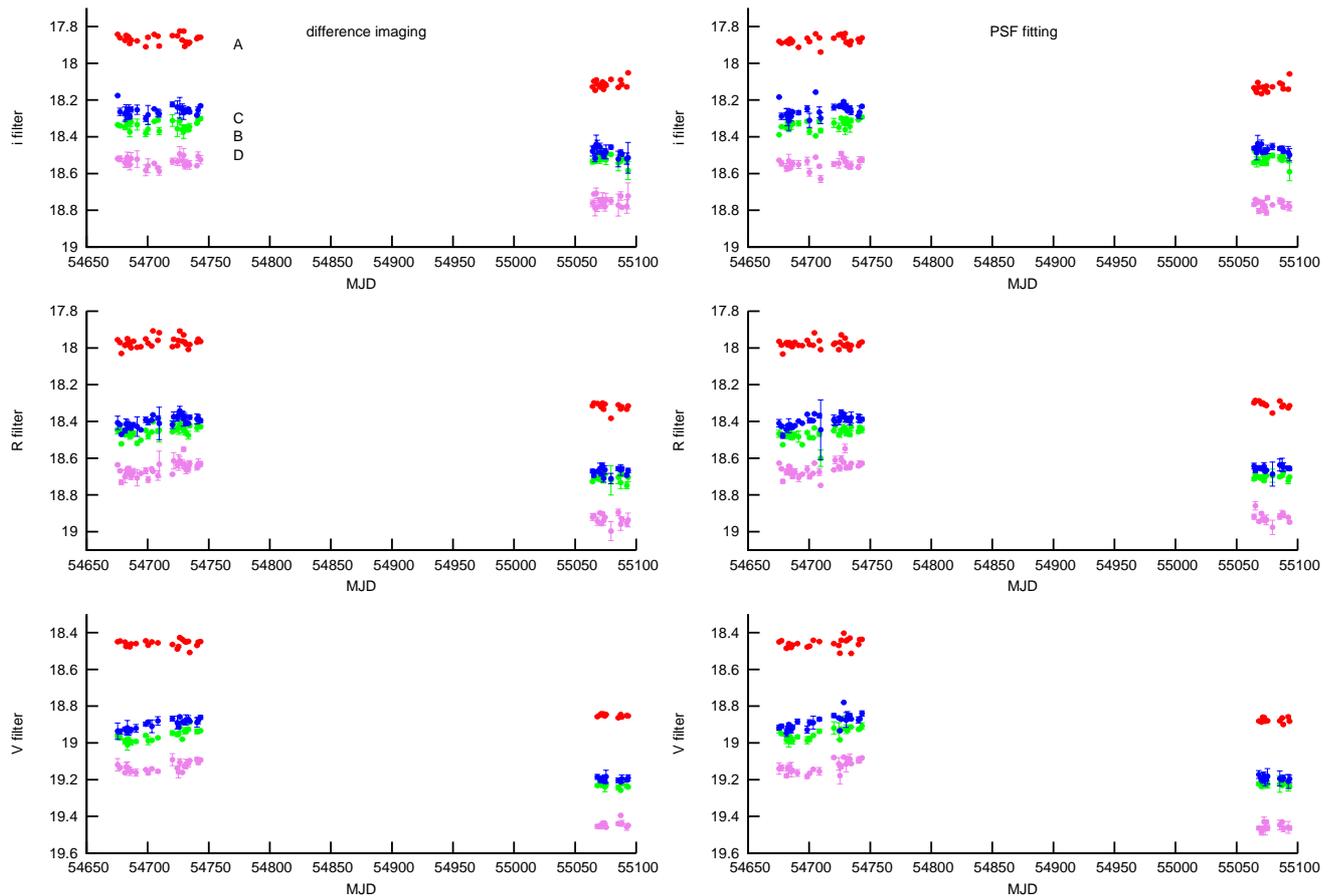}
  \caption{Light curves of the four lensed components of \he. The
    different graphs illustrate the photometry in the $i$, $R$ and $V$
    bands, calculated using the difference imaging technique (left)
    and the PSF fitting technique (right). The error bars correspond
    to the magnitude rms ($1 \sigma$) of each night of observation.}
  \label{fig:curves}
\end{figure*}

\noindent For the calculation of the ratio between the fluxes of two
selected stars we applied the PSF fitting method that was also
subsequently used on the gravitational lens system itself. While using
this method, we superposed the image of the first star over the fixed
image of the second star as shown in Fig.~\ref{fig:grid}.  The red and
black grids represent the images of the first and the second star,
respectively.

The big dots correspond to the knots for which the counts are
known. The values $p(x,y)$ and $S$ correspond to the counts for each
knot of the first and the second star, respectively.  To perform a
precise fitting we needed to superpose these images with an accuracy
better than 1 pixel.  When this was done, the knots of both images did
not perfectly coincide with each other as shown in
Fig.~\ref{fig:grid}.  Therefore we had to calculate intermediate
values, for example $p(x,y)$, which we could compare with the value $S$
at the same point.  For that we used bicubic interpolation
\citep{numrec}. The intermediate value $p(x,y)$ is expressed by the
polynomial
\begin{equation}
  p(x,y)=\sum_{i=0}^{3} \sum_{j=0}^{3}a_{i,j}x^{i}y^{j} . 
\end{equation} 
To derive the values of the 16 coefficients $a_{i,j}$ we resolved this
set of equations for 16 knots around the considered point
$p(x,y)$. They are shown as red dots in Fig.~\ref{fig:grid}. With this
coefficient matrix $a_{i,j}$ we could derive $p(x,y)$ at any
point. For the fitting of both images we minimized the quantity
\begin{equation}
  \Delta (A, \Delta x, \Delta y)= \sum_{k} (A\cdot p_{k}(x+\Delta x, y+\Delta y)-S_{k})^{2},
\end{equation} 
which is the sum over all knots $k$ of the first star image; $A$ is
the ratio between the total flux of the second and the first star,
$\Delta x$ and $\Delta y$ are the relative shifts (fractions of a
pixel) between the two superposed images.  Before the first iteration,
we set the three parameters $A$, $\Delta x$ and $\Delta y$ within
reasonable ranges. 
\begin{table}[b]
  \caption{ Average V magnitude differences between the two epochs for the four stars near \he. }
\label{tab:reference}
  \centering
  \begin{tabular}{ccc}
    \hline\hline 
    Pair   &  $< V_{2008} >$    &  $< V_{2009} >$    \\
    \hline 
    $<$S$-$R$>$  &  $\phantom{-}1.78 \pm 0.03$ & $ \phantom{-}1.77 \pm 0.02$ \\ 
    $<$S$-$T$>$  &  $\phantom{-}2.04 \pm 0.07$ & $ \phantom{-}1.95 \pm 0.07$ \\ 
    $<$T$-$R$>$  &            $-0.41 \pm 0.06$ &            $-0.40 \pm 0.04$ \\ 
    $<$U$-$S$>$  &            $-2.36 \pm 0.04$ &            $-2.42 \pm 0.03$ \\  
    $<$U$-$R$>$  &            $-0.58 \pm 0.04$ &            $-0.65 \pm 0.03$ \\ 
    $<$U$-$T$>$  &            $-0.34 \pm 0.05$ &            $-0.31 \pm 0.04$ \\ 
    \hline
  \end{tabular}
\end{table}

\begin{figure*}[t]
  \centering
  \includegraphics[width=17.4cm]{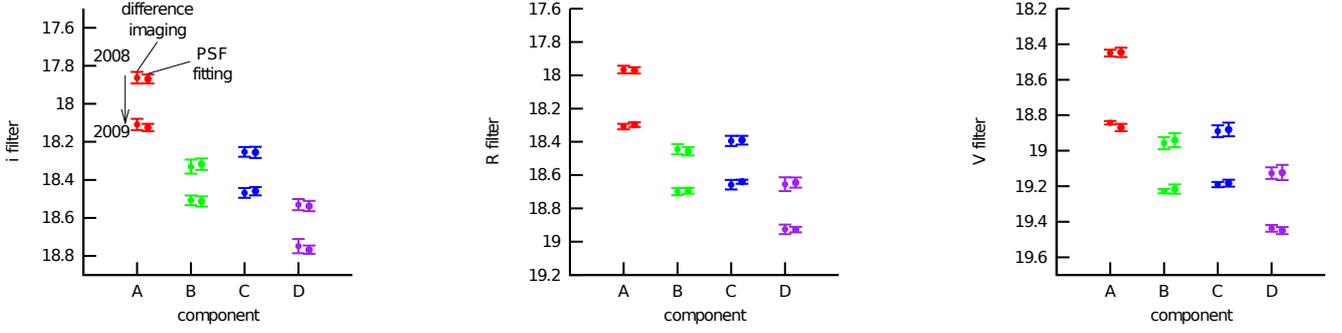}
  \caption{Average magnitude of each component for the 2008 season
    (upper symbols) and 2009 (lower symbols), calculated with the difference imaging technique (left) and the PSF fitting method (right).
The error bars correspond to the magnitude rms ($1 \sigma$) during each epoch of
observation.}
  \label{fig:media}
\end{figure*}

\begin{table*}
  \caption{$VRi$ average magnitudes of the four lensed components of \he during the 2008 and 2009 seasons.}
\label{tab:avecurves}
  \centering
  \begin{tabular}{clcccc}
    \hline\hline 
    2008  & technique &   A                     & B                          & C                            & D \\
    \hline
    $i$ & difference imaging  &  $17.86 \pm 0.03 $ &  $  18.33 \pm 0.04 $ &   $   18.25 \pm 0.03 $ &   $   18.53 \pm 0.03 $ \\
    $i$ & PSF fitting         &  $17.87 \pm 0.02 $ &  $  18.32 \pm 0.03 $ &   $   18.26 \pm 0.03 $ &   $   18.54 \pm 0.03 $ \\
    $R$ & difference imaging  &  $17.97 \pm 0.02 $ &  $  18.46 \pm 0.02 $ &   $   18.39 \pm 0.03 $ &   $   18.64 \pm 0.03 $ \\
    $R$ & PSF fitting         &  $17.97 \pm 0.02 $ &  $  18.44 \pm 0.03 $ &   $   18.39 \pm 0.03 $ &   $   18.65 \pm 0.04 $ \\
    $V$ & difference imaging  &  $18.45 \pm 0.03 $ &  $  18.94 \pm 0.04 $ &   $   18.88 \pm 0.04 $ &   $   19.12 \pm 0.04 $ \\
    $V$ & PSF fitting         &  $18.45 \pm 0.02 $ &  $  18.96 \pm 0.03 $ &   $   18.89 \pm 0.03 $ &   $   19.13 \pm 0.03 $ \\
    \hline
    2009  & technique &   A                   & B                   & C                     & D \\
    \hline
    $i$ & difference imaging  &  $18.11 \pm 0.03 $ &  $  18.51 \pm 0.02 $ &   $   18.47 \pm 0.03 $ &   $   18.75 \pm 0.04 $ \\
    $i$ & PSF fitting         &  $18.12 \pm 0.02 $ &  $  18.51 \pm 0.03 $ &   $   18.46 \pm 0.02 $ &   $   18.77 \pm 0.02 $ \\
    $R$ & difference imaging  &  $18.30 \pm 0.01 $ &  $  18.69 \pm 0.02 $ &   $   18.64 \pm 0.01 $ &   $   18.93 \pm 0.02 $ \\
    $R$ & PSF fitting         &  $18.31 \pm 0.02 $ &  $  18.70 \pm 0.02 $ &   $   18.66 \pm 0.03 $ &   $   18.93 \pm 0.03 $ \\
    $V$ & difference imaging  &  $18.87 \pm 0.02 $ &  $  19.22 \pm 0.03 $ &   $   19.18 \pm 0.02 $ &   $   19.45 \pm 0.02 $ \\
    $V$ & PSF fitting         &  $18.84 \pm 0.01 $ &  $  19.23 \pm 0.01 $ &   $   19.19 \pm 0.02 $ &   $   19.44 \pm 0.02 $ \\
    \hline
  \end{tabular}
\end{table*}

\noindent We derived the light curves for the reference stars as
magnitude differences and calculated the average difference and
standard deviations for the two epochs (see
Table~\ref{tab:reference}). Obviously the stellar pairs with the star
``R'' in Table~\ref{tab:reference} show on average the smallest
differences between the two epochs. Moreover the light curves of the
``R'' star shows on average the least standard deviation $\sigma$ (see
Table~\ref{tab:reference}), and we may reasonably assume that
star ``R'' is the most stable reference star between the two seasons.
Therefore we chose star ``R'' as the reference for all subsequent
photometric zero point determinations.

\noindent The magnitude of the reference star was taken from the
USNO-B1.0 catalog for the $i$ and $R$ filters (16.27 and 16.33
respectively), and from the NOMAD1 catalog for the $V$ filter (17.04).

\noindent The light curves for the four components of the gravitational
lens system were then calculated with two independent methods treated
below: difference imaging and PSF fitting.

\subsection{Difference imaging method}

The aim of the difference imaging technique is to subtract from each
image of our field (indicated as ``frame'' in the following) one image
of the same field (called ``reference frame'') taken at a different
time under the best seeing conditions. This operation produces a set
of subtracted frames where only the relative flux variations between
the two images (generic frame and reference frame) are
visible. Performing aperture photometry on these subtracted frames,
and in particular at the positions of the lensed QSO components, we
derived the light curves of the four lensed components.

\begin{figure*}[t]
  \centering
  \includegraphics[width=17.4cm]{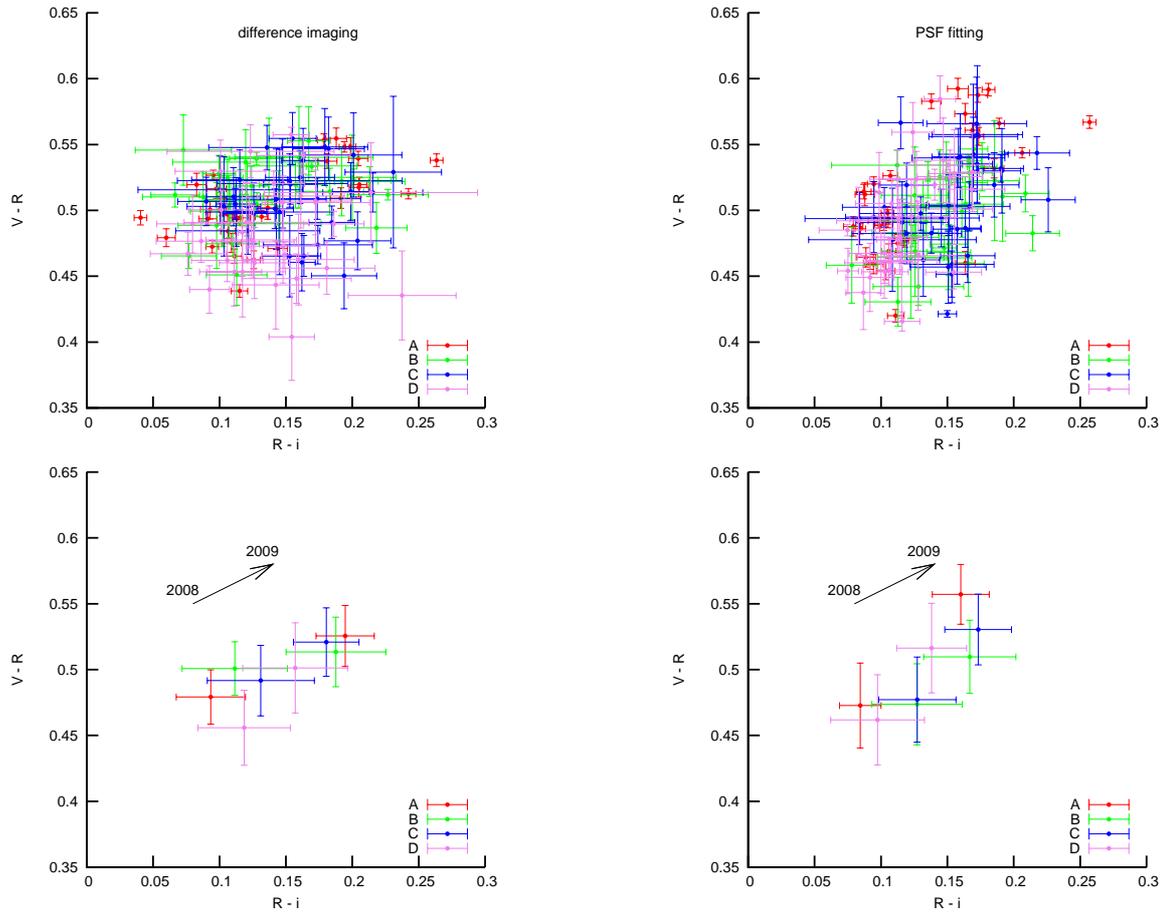}
  \caption{Color-color diagrams for the four lensed components. The left
    column shows the $V-R$ vs. $R-i$ diagram calculated with the
    difference imaging technique: all observations (upper panel) and
    average of the observations for each season (lower panel). The
    right column shows the same diagrams calculated with the PSF
    fitting technique.}
  \label{fig:color}
\end{figure*}

\begin{table*}
  \caption{Averages and error bars (sigma) characterizing the $V-R$ and $R-i$ color indices of each component for the 2008 and 2009 seasons.}
  \label{tab:avecolor}
  \centering
  \begin{tabular}{clcccc}
    \hline\hline 
    2008  & technique & A                     & B                          & C                            & D \\
    \hline
    $V-R$ & difference imaging &    $0.47 \pm 0.03 $ &  $ 0.47 \pm 0.03 $ &   $  0.48 \pm 0.03 $ &   $  0.46 \pm 0.03 $ \\
    $V-R$ & PSF fitting        &    $0.48 \pm 0.02 $ &  $ 0.50 \pm 0.02 $ &   $  0.49 \pm 0.03 $ &   $  0.46 \pm 0.03 $ \\
    $R-i$ & difference imaging &    $0.08 \pm 0.02 $ &  $ 0.13 \pm 0.03 $ &   $  0.13 \pm 0.03 $ &   $  0.10 \pm 0.04 $ \\
    $R-i$ & PSF fitting        &    $0.09 \pm 0.03 $ &  $ 0.11 \pm 0.04 $ &   $  0.13 \pm 0.04 $ &   $  0.12 \pm 0.03 $ \\
    \hline
    2009  & technique &  A                   & B                  & C                  & D \\
    \hline
    $V-R$ & difference imaging &    $0.56 \pm 0.02 $ &  $ 0.51 \pm 0.03 $ &   $  0.53 \pm 0.03 $ &   $  0.52 \pm 0.03 $ \\
    $V-R$ & PSF fitting        &    $0.53 \pm 0.02 $ &  $ 0.51 \pm 0.03 $ &   $  0.52 \pm 0.03 $ &   $  0.50 \pm 0.03 $ \\
    $R-i$ & difference imaging &    $0.16 \pm 0.02 $ &  $ 0.17 \pm 0.03 $ &   $  0.17 \pm 0.03 $ &   $  0.14 \pm 0.03 $ \\
    $R-i$ & PSF fitting        &    $0.19 \pm 0.02 $ &  $ 0.19 \pm 0.04 $ &   $  0.18 \pm 0.02 $ &   $  0.16 \pm 0.04 $ \\
\hline
  \end{tabular}
\end{table*}

\noindent However, differences in seeing, focus, and guiding precision
between frames collected at different times may produce variations in
the shape of the PSF: trying the subtraction without additional
operations would produce high residuals caused by potential PSF slope
variations. Several methods have been developed to force the PSF of
the images to match \citep{alard99,alard00}. These methods are
particularly useful in crowded fields such as the galactic bulge, but
are less succesful in sparse fields. In this paper we adopt the method
proposed by \cite{philips93} which was already successfully applied by
\cite{nakos05}and is based on FFT (Fast Fourier Transform).

\noindent If $r$ is the reference frame and $f$ a generic frame, then
$
  f = r	\otimes k
  \label{eq:rfk}
$,
where $k$ is the convolution kernel describing the differences between
the PSF, which are unknown, and $\otimes$ indicates the convolution product. 
In the Fourier space, the previous equation can be noted as
$
  F = R K
  \label{eq:RFK}
$,
where $F$, $R$ and $K$ represents the Fourier transform ($\mathscr{F}$)
of the generic frame $f$, the reference frame $r$, and the convolution
kernel $k$. Then 
$ k = \mathscr{F}^{-1}(F/R)
  \label{eq:findingk}
$.  

\noindent \cite{philips93} considers the limits of this technique and the
solutions adopted to avoid problems with background and high-frequency
noise.

\noindent We normalized the frames in flux by fixing the magnitude of
the reference star in each filter with the values of the catalogs
mentioned above. Then we used the difference imaging method on the
normalized frames. With this technique, all the not variable objects
in the field disappear, which allows us automatically to suppress the
contribution of the lensing galaxy, which is an extended and
photometrically constant object.

\noindent To obtain the light curves, we performed aperture photometry
of the residuals, using the positions of the selected reference star and of the
lensed components previously derived. But we found for the 2009
images a weak linear dependence between the magnitude and the seeing, 
which we removed after calibrating this effect.

\noindent The procedure described in this paragraph is based on a code
developed by our team, written in Python.

\noindent The results for the three filters are shown in the left column
of Fig.~\ref{fig:curves}. The error bars correspond to the magnitude
rms ($1 \sigma$) of each night of observation.

\subsection{PSF fitting method}

\noindent We also decided to calculate the light curves using PSF
fitting as an independent method, which we previously employed to
determine the most adequate reference star.

\noindent For the fitting of the lens system we used the image of star
``R'' as the PSF reference. Then we fitted each frame with five
adjustable PSF for the four lensed quasar images and the lensing
galaxy, taking the relative astrometric coordinates between the
components from \cite{kochanek06}. Note that the faint lensing galaxy
is barely resolved on direct HST CCD frames
\citep{morgan05}\footnote{HST program 9744 }. 
Therefore, it is legitimate to model it with the PSF of our
ground-based observations.

\noindent In this way we had seven free parameters: $ \Delta x$,
$\Delta y$ (coordinates of the gravitational lens system with respect
to the reference star ``R''), and the central fluxes of the five
components.

\noindent After minimization of the squared differences between the
fluxes of the real lens system image and the simulated image with the
five PSF, we derived the seven best-fitting parameters. We used
bicubic interpolation for the superposition of the CCD frames to
achieve results better than 1 pixel according to our description in
Sect.~\ref{sec:red}.  To construct the light curves, we calculated the
flux ratios between each component and the reference star ``R''. The
procedure described in this paragraph is based on an Object Pascal
code developed by our team.

\noindent The results for the three filters are shown in the right column
of Fig.~\ref{fig:curves}. The error bars correspond to the magnitude
rms ($1 \sigma$) of each night of observation.
\begin{figure*}[t]
  \centering
  \includegraphics[width=17.6cm]{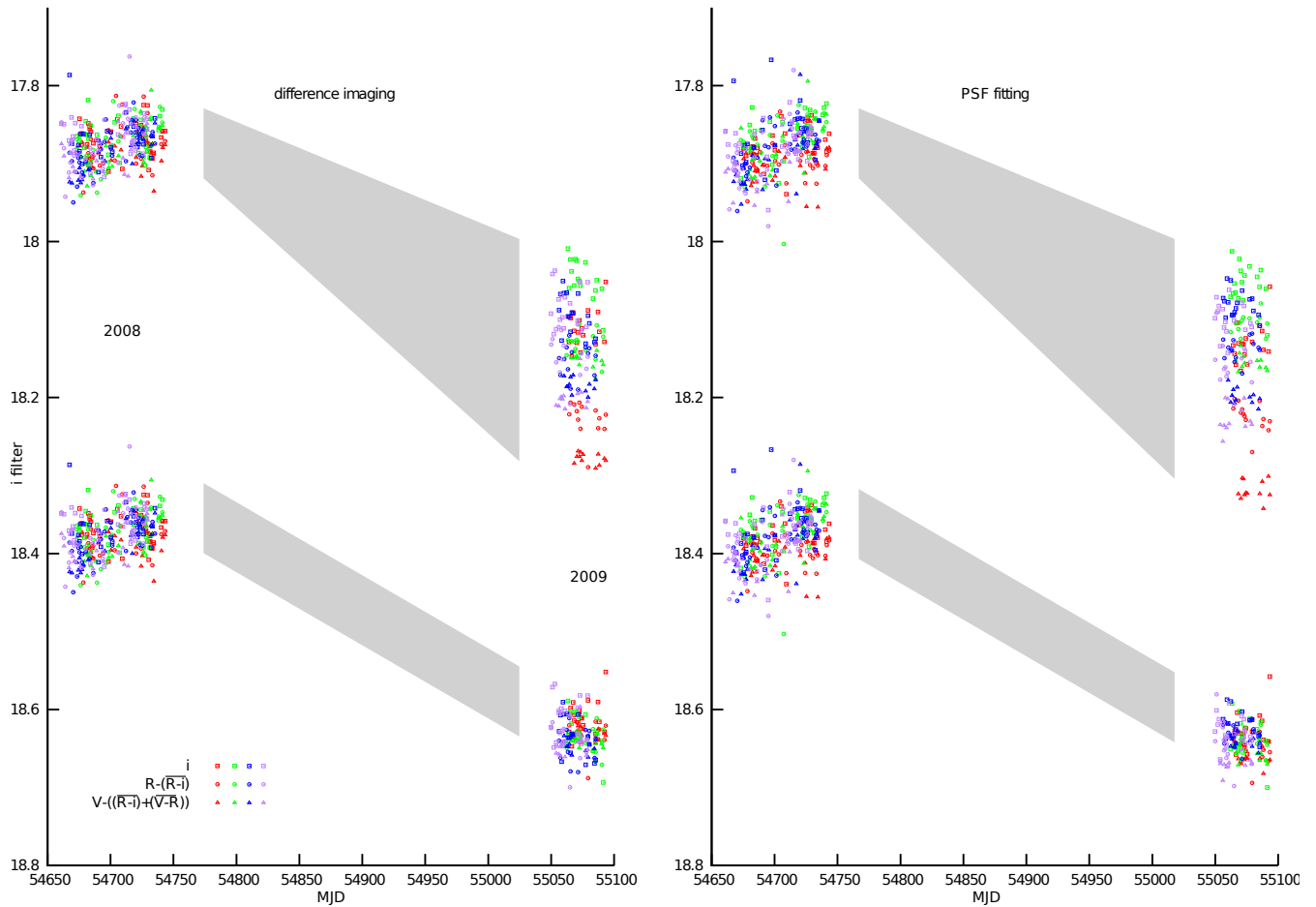}
  \caption{ Upper light curve: ``global $i$'' light curve of the four
    lensed components of \he obtained after subtracting the average
    2008 $i$ difference in magnitude from ``B'', ``C'', and ``D'' with
    respect to component ``A''; then corrected for the time delays
    provided by \cite{kochanek06} and finally subtracting from the $R$
    and the $V$ light curves the average 2008 $R-i$ and $(R-i)+(V-R)$
    color indices, respectively.  Lower light curve: ``global $i$''
    light curve obtained repeating the same procedure as for the upper
    one, but correcting for the average 2008 magnitude and color
    indices concerning the 2008 data, and for the 2009 average
    magnitude and color indices concerning the 2009 data. The left
    panel shows the results obtained with the difference imaging
    technique, while the right panel shows the results obtained with
    the PSF fitting technique. The gray quadrilaterals help to connect
    the two epochs of observation. The lower light curves are arbitrarily shifted in magnitude.}
  \label{fig:shifted}
\end{figure*}

\section{Results}
\label{sec:res}

Both methods show a significant decrease in flux of the four lensed
components between the 2008
and 2009 seasons. The estimated amounts of the decrease are coherent between
the two methods (see Fig.~\ref{fig:media}).

\noindent In order to estimate this decrease, we measured the mean and
the $\sigma$ for each component in each filter.  The average values
for the magnitudes and rms for each component, each filter and the
2008 and 2009 seasons, are reported in
Table~\ref{tab:avecurves}. Specifically, all the four components show
a decrease by $\approx$ 0.2--0.4 magnitudes in each of the filters,
although we notice a slightly larger amplitude for component ``A'' in
the $V$ band.

\noindent The corresponding values expressed in sigma units show a
shift between $\approx 11.3\sigma$ and $\approx 13.7\sigma$ in the $V$
band, except for component ``A'', which shows a decrease by $\approx
26.3\sigma$.  In the $R$ and $i$ filters, the shift is between
$\approx 6.5\sigma$ and $7.5\sigma$, except for component ``A''
($\approx 15.0\sigma$ and $\approx 9.8\sigma$ respectively) and
component ``C'' in the $R$ band ($\approx 12.0\sigma$). 

For the fraction of nights when the object was observed in all $VRi$
filters, we were able to build the color-color diagram ($V-R$
vs. $R-i$) for the four components. The results are shown in
Fig.~\ref{fig:color}. With the same technique used to estimate the
decrease in flux, we also found a significant increase ($\approx$
0.05--0.015) for the color indices $V-R$ and $R-i$ between the two
observing seasons. The details are given in
Table~\ref{tab:avecolor}. In particular, component ``A'' shows the
largest shift in color.

\noindent The corresponding values expressed in sigma units show a
shift in color between $\approx 1.3\sigma$ and $\approx 2.0\sigma$ for
the color indices $V-R$ and $R-i$, except for component ``A''
($3.40\sigma$ in $V-R$ and $3.1\sigma$ in $R-i$).

\noindent Given the short time delays (\cite{kochanek06}) and because
we expect microlensing to lead to uncorrelated flux variations between
the four lensed components, these results support the assumption that
the observed magnitude and color variations are very likely caused by
intrinsic variations of the QSO, while the lensed ``A'' component is
probably also affected by microlensing.

As a complementary approach, we decided to construct two ``global
$i$'' light curves to better understand the nature of these different
variations in flux and color index.

\noindent To construct the first one, we superposed the light curves
of the four components by subtracting from ``B'', ``C'', and ``D''
their average 2008 difference in magnitude with respect to the ``A''
component, and we corrected the data for the time delays provided by
\cite{kochanek06} (i.e, we applied to the components a shift in the
MJD corresponding to the time delays). Then, we superposed the
obtained curves (one for each filter) by subtracting from the $R$ and
the $V$ light curves the average 2008 $R-i$ and $(R-i)+(V-R)$ color
indices, respectively. The goal of this first ``global $i$'' light
curve is to visualize how the spreads in magnitude and colors evolve
between the epochs. The results are shown in Fig.~\ref{fig:shifted}
(upper light curves).

\noindent To construct the second ``global $i$'' light curve, we
repeated the same procedure, but subtracted from the 2008 data of
``B'', ``C'' and ``D'' the average 2008 difference in magnitude with
respect to component ``A'', and from 2009 data the corresponding
average 2009 difference in magnitude. Similarly, we subtracted from the
$R$ and the $V$ light curves the average 2008 and 2009 $R-i$ and
$(R-i)+(V-R)$ color indices, respectively.  The results are shown in
Fig.~\ref{fig:shifted} (lower curves).

After these superpositions, we observe in 2008 a scatter in the data
(see Fig.~\ref{fig:shifted}) significantly larger than that of the
individual light curves (see Fig.~\ref{fig:curves}); and in general a
difference in scatter between the two epochs that we attribute to 
intrinsic variations of the quasar both in magnitude and color.

\noindent We also observe a slight brightening of the lensed quasar in
2008, followed by a significant decrease in 2009. Our observations
corroborate those recently reported by \cite{courbin10}.
Referring to the PSF-only fitting method, we were also able to
estimate the magnitude of the lensing galaxy as
$19.87\pm0.10$ in the $i$ band;
$20.47\pm0.13$ in the $R$ band; and
$21.89\pm0.24$ in the $V$ band.
%
%
Furthermore, no significant changes in the magnitude of the lensing
galaxy were observed between the two epochs. Our results for the $V$
and  $i$ filters are coherent with those of \cite{wisotzki02} and
\cite{morgan05}.

\noindent Proceeding in the same way, we did not find any evident
color-shift for the lensing galaxy: we find a value of $1.47\pm0.33$
and $1.36\pm0.21$ for the $V-R$ color index in 2008 and 2009,
respectively; and $0.59\pm0.17$ and $0.61\pm0.14$ for the $R-i$ color
index.

\noindent The stability of the flux of the lensing galaxy over the two
epochs enforces the validity of our PSF fitting software.  We find
that appreciably larger photometric error bars derived for the lensing
galaxy are essentially caused by its faintness ($V\approx21.9$)
compared with the four lensed components ($V \in [18.4,19.5]$).

\section{Conclusion}
\label{sec:conc}
Systematic multi-spectral band photometry of the quadruply imaged
quasar \he, carried out during two seasons (2008 and 2009), shows a
significant decrease in flux of the four lensed components between the
two epochs.

\noindent The drop in flux observed for the four lensed components between
2008 and 2009 is very likely caused by a change in the intrinsic
luminosity of the quasar. This corroborates the previous studies of
the \he gravitational lens system by \cite{kochanek06} and
\cite{courbin10}. 
\noindent Concerning the color variations, the intrinsic reddening of
a quasar when it becomes fainter in luminosity is an effect already
observed in previous studies \citep{pereyra06}, and the same trend
probably accounts for the similar observed changes in the colors of
the four lensed components of \he. 

\noindent This hypothesis is enforced if we suppose that the intrinsic
photometric quasar variations in the different colors are not
synchronized, which provides an explanation for the differences in
scatter between the epochs (see Fig.~\ref{fig:shifted}). Microlensing
probably provides the additional effect necessary to account for
the higher flux variation observed for the ``A'' component.

\noindent The presented observations also show that a well sampled
multi-band photometry can help in distinguishing the nature of the
variability of multiply imaged objects, in particular gravitationally
lensed quasars. 

We suggest  to couple this technique in the future with
integral field spectroscopy to provide an additional way for an even
more detailed investigation of the observed phenomena.

\begin{acknowledgements}
  The research was supported by ARC -- Action de recherche concert\'ee
  (Communaut\'e Fran\c caise de Belgique -- Acad\'emie
  Wallonie-Europe). AE is the beneficiary of a fellowship granted by
  the Belgian Federal Science Policy Office. Astronomical research at
  Armagh Observatory is funded by the Department of Culture, Arts \&
  Leisure (DCAL), Northern Ireland, UK. Operation of the Danish
  $1.54\meter$ telescope is supported by the Danish National Science
  Research Council (FNU). We wish to thank the anonymous referee for
  the remarks and the suggestions.
\end{acknowledgements}

\bibliography{biblio,bibliorf} 
\bibliographystyle{aa} 

\Online

\onecolumn

 \begin{landscape}
   \begin{table}
     \centering
     \caption{\label{istars} Light curves in the $i$ filter}
     \begin{tabular}{cccccccccc}
       \hline
       \hline
       \multicolumn{5}{c}{Difference imaging}  &  \multicolumn{5}{c}{PSF fitting}  \\
       MJD & A & B & C & D                       &  MJD & A & B & C & D \\
       \hline
       54675.4 & $17.84\pm0.01$ & $18.33\pm0.01$ & $18.18\pm0.01$ & $18.52\pm0.01$ & 	54675.4 & $17.88\pm0.01$ & $18.39\pm0.01$ & $18.18\pm0.01$ & $18.53\pm0.01$  \\
54677.4 & $17.86\pm0.01$ & $18.34\pm0.01$ & $18.26\pm0.02$ & $18.52\pm0.02$ & 	54677.4 & $17.89\pm0.01$ & $18.35\pm0.01$ & $18.29\pm0.02$ & $18.55\pm0.01$  \\
54681.4 & $17.88\pm0.01$ & $18.35\pm0.01$ & $18.28\pm0.04$ & $18.54\pm0.02$ & 	54681.4 & $17.88\pm0.01$ & $18.35\pm0.01$ & $18.28\pm0.03$ & $18.57\pm0.02$  \\
54682.4 & $17.85\pm0.01$ & $18.32\pm0.02$ & $18.27\pm0.04$ & $18.54\pm0.02$ & 	54682.4 & $17.88\pm0.01$ & $18.32\pm0.02$ & $18.29\pm0.05$ & $18.57\pm0.01$  \\
54683.4 & $17.85\pm0.01$ & $18.33\pm0.03$ & $18.27\pm0.03$ & $18.51\pm0.03$ & 	54683.4 & $17.89\pm0.01$ & $18.35\pm0.02$ & $18.31\pm0.05$ & $18.53\pm0.04$  \\
54684.4 & $17.86\pm0.01$ & $18.29\pm0.04$ & $18.25\pm0.03$ & $18.54\pm0.01$ & 	54684.4 & $17.87\pm0.01$ & $18.30\pm0.03$ & $18.27\pm0.02$ & $18.57\pm0.01$  \\
54685.4 & $17.89\pm0.01$ & $18.37\pm0.03$ & $18.29\pm0.01$ & $18.55\pm0.03$ & 	54685.4 & $17.89\pm0.01$ & $18.34\pm0.02$ & $18.29\pm0.03$ & $18.55\pm0.02$  \\
54686.4 & $17.87\pm0.01$ & $18.32\pm0.02$ & $18.25\pm0.01$ & $18.52\pm0.03$ & 	54686.4 & $17.88\pm0.01$ & $18.33\pm0.01$ & $18.26\pm0.01$ & $18.54\pm0.02$  \\
54691.4 & $17.88\pm0.01$ & $18.33\pm0.03$ & $18.25\pm0.03$ & $18.52\pm0.04$ & 	54691.4 & $17.91\pm0.01$ & $18.33\pm0.01$ & $18.27\pm0.01$ & $18.55\pm0.02$  \\
54698.4 & $17.91\pm0.01$ & $18.38\pm0.02$ & $18.30\pm0.02$ & $18.58\pm0.03$ & 	54698.4 & $17.86\pm0.01$ & $18.32\pm0.01$ & $18.25\pm0.02$ & $18.53\pm0.02$  \\
54700.3 & $17.86\pm0.01$ & $18.36\pm0.03$ & $18.28\pm0.05$ & $18.56\pm0.04$ & 	54700.3 & $17.88\pm0.01$ & $18.37\pm0.02$ & $18.31\pm0.04$ & $18.59\pm0.02$  \\
54705.4 & $17.84\pm0.01$ & $18.32\pm0.01$ & $18.25\pm0.01$ & $18.55\pm0.01$ & 	54705.4 & $17.84\pm0.01$ & $18.40\pm0.01$ & $18.16\pm0.01$ & $18.51\pm0.01$  \\
54708.3 & $17.85\pm0.01$ & $18.31\pm0.01$ & $18.27\pm0.02$ & $18.57\pm0.01$ & 	54708.3 & $17.86\pm0.01$ & $18.32\pm0.01$ & $18.27\pm0.03$ & $18.56\pm0.01$  \\
54709.4 & $17.91\pm0.01$ & $18.37\pm0.02$ & $18.27\pm0.02$ & $18.59\pm0.02$ & 	54709.4 & $17.94\pm0.01$ & $18.37\pm0.01$ & $18.30\pm0.03$ & $18.63\pm0.02$  \\
54720.3 & $17.85\pm0.01$ & $18.31\pm0.03$ & $18.22\pm0.01$ & $18.53\pm0.02$ & 	54720.3 & $17.86\pm0.01$ & $18.32\pm0.02$ & $18.24\pm0.02$ & $18.55\pm0.01$  \\
54724.3 & $17.85\pm0.01$ & $18.36\pm0.04$ & $18.24\pm0.03$ & $18.54\pm0.02$ & 	54724.3 & $17.85\pm0.01$ & $18.35\pm0.02$ & $18.23\pm0.01$ & $18.54\pm0.03$  \\
54726.3 & $17.82\pm0.01$ & $18.31\pm0.02$ & $18.24\pm0.06$ & $18.49\pm0.04$ & 	54726.3 & $17.84\pm0.01$ & $18.30\pm0.01$ & $18.23\pm0.01$ & $18.49\pm0.01$  \\
54728.3 & $17.87\pm0.01$ & $18.36\pm0.02$ & $18.25\pm0.03$ & $18.53\pm0.02$ & 	54728.3 & $17.86\pm0.01$ & $18.32\pm0.02$ & $18.21\pm0.01$ & $18.52\pm0.01$  \\
54729.4 & $17.83\pm0.01$ & $18.37\pm0.04$ & $18.27\pm0.04$ & $18.50\pm0.04$ & 	54729.4 & $17.84\pm0.01$ & $18.36\pm0.03$ & $18.24\pm0.01$ & $18.53\pm0.01$  \\
54730.3 & $17.91\pm0.01$ & $18.35\pm0.01$ & $18.26\pm0.01$ & $18.56\pm0.02$ & 	54730.3 & $17.89\pm0.01$ & $18.30\pm0.01$ & $18.23\pm0.01$ & $18.54\pm0.03$  \\
54731.4 & $17.88\pm0.01$ & $18.35\pm0.02$ & $18.26\pm0.02$ & $18.54\pm0.02$ & 	54731.4 & $17.89\pm0.01$ & $18.33\pm0.01$ & $18.26\pm0.01$ & $18.56\pm0.01$  \\
54733.3 & $17.89\pm0.01$ & $18.36\pm0.01$ & $18.25\pm0.01$ & $18.55\pm0.01$ & 	54733.3 & $17.90\pm0.01$ & $18.34\pm0.03$ & $18.24\pm0.01$ & $18.56\pm0.01$  \\
54734.3 & $17.89\pm0.01$ & $18.35\pm0.02$ & $18.27\pm0.01$ & $18.55\pm0.02$ & 	54734.3 & $17.88\pm0.01$ & $18.31\pm0.01$ & $18.26\pm0.02$ & $18.55\pm0.02$  \\
54740.3 & $17.87\pm0.01$ & $18.33\pm0.01$ & $18.28\pm0.01$ & $18.56\pm0.01$ & 	54740.3 & $17.87\pm0.01$ & $18.31\pm0.01$ & $18.29\pm0.01$ & $18.57\pm0.01$  \\
54741.3 & $17.86\pm0.01$ & $18.31\pm0.01$ & $18.25\pm0.02$ & $18.51\pm0.02$ & 	54741.3 & $17.88\pm0.01$ & $18.30\pm0.01$ & $18.27\pm0.03$ & $18.53\pm0.02$  \\
54743.3 & $17.86\pm0.01$ & $18.30\pm0.01$ & $18.23\pm0.01$ & $18.53\pm0.02$ & 	54743.3 & $17.86\pm0.01$ & $18.29\pm0.01$ & $18.23\pm0.01$ & $18.52\pm0.02$  \\ 
55064.3 & $18.13\pm0.01$ & $18.54\pm0.01$ & $18.48\pm0.03$ & $18.76\pm0.02$ & 	55064.3 & $18.13\pm0.01$ & $18.54\pm0.01$ & $18.46\pm0.01$ & $18.77\pm0.01$  \\
55065.4 & $18.10\pm0.01$ & $18.48\pm0.01$ & $18.46\pm0.01$ & $18.71\pm0.01$ & 	55065.4 & $18.14\pm0.01$ & $18.48\pm0.01$ & $18.47\pm0.01$ & $18.74\pm0.01$  \\
55066.4 & $18.15\pm0.01$ & $18.53\pm0.02$ & $18.52\pm0.02$ & $18.79\pm0.04$ & 	55066.4 & $18.16\pm0.01$ & $18.54\pm0.03$ & $18.49\pm0.04$ & $18.76\pm0.01$  \\
55067.4 & $18.09\pm0.01$ & $18.49\pm0.04$ & $18.44\pm0.05$ & $18.71\pm0.03$ & 	55067.4 & $18.10\pm0.01$ & $18.52\pm0.01$ & $18.44\pm0.04$ & $18.75\pm0.01$  \\
55068.3 & $18.11\pm0.01$ & $18.51\pm0.03$ & $18.46\pm0.04$ & $18.78\pm0.02$ & 	55068.3 & $18.13\pm0.01$ & $18.54\pm0.01$ & $18.47\pm0.01$ & $18.80\pm0.01$  \\
55070.4 & $18.13\pm0.01$ & $18.53\pm0.02$ & $18.46\pm0.02$ & $18.74\pm0.01$ & 	55070.4 & $18.17\pm0.01$ & $18.54\pm0.01$ & $18.44\pm0.02$ & $18.76\pm0.01$  \\
55071.4 & $18.12\pm0.01$ & $18.49\pm0.03$ & $18.49\pm0.01$ & $18.77\pm0.02$ & 	55071.4 & $18.13\pm0.01$ & $18.49\pm0.02$ & $18.48\pm0.01$ & $18.78\pm0.03$  \\
55072.4 & $18.10\pm0.01$ & $18.49\pm0.03$ & $18.49\pm0.01$ & $18.74\pm0.03$ & 	55072.4 & $18.13\pm0.01$ & $18.51\pm0.02$ & $18.48\pm0.01$ & $18.77\pm0.03$  \\
55073.4 & $18.14\pm0.01$ & $18.52\pm0.01$ & $18.51\pm0.01$ & $18.77\pm0.01$ & 	55073.4 & $18.14\pm0.01$ & $18.52\pm0.01$ & $18.48\pm0.01$ & $18.79\pm0.02$  \\
55074.4 & $18.11\pm0.01$ & $18.52\pm0.01$ & $18.48\pm0.02$ & $18.78\pm0.02$ & 	55074.4 & $18.12\pm0.01$ & $18.55\pm0.01$ & $18.48\pm0.02$ & $18.81\pm0.01$  \\
55075.4 & $18.12\pm0.01$ & $18.53\pm0.02$ & $18.48\pm0.04$ & $18.74\pm0.03$ & 	55075.4 & $18.16\pm0.01$ & $18.51\pm0.01$ & $18.47\pm0.02$ & $18.73\pm0.01$  \\
55079.4 & $18.09\pm0.01$ & $18.50\pm0.01$ & $18.46\pm0.01$ & $18.75\pm0.02$ & 	55079.4 & $18.13\pm0.01$ & $18.50\pm0.01$ & $18.45\pm0.02$ & $18.77\pm0.01$  \\
55085.4 & $18.13\pm0.01$ & $18.54\pm0.04$ & $18.52\pm0.04$ & $18.77\pm0.06$ & 	55085.4 & $18.11\pm0.01$ & $18.52\pm0.02$ & $18.46\pm0.01$ & $18.75\pm0.02$  \\
55087.3 & $18.09\pm0.01$ & $18.53\pm0.01$ & $18.49\pm0.01$ & $18.72\pm0.02$ & 	55087.3 & $18.11\pm0.01$ & $18.53\pm0.01$ & $18.46\pm0.01$ & $18.75\pm0.01$  \\
55088.3 & $18.12\pm0.01$ & $18.52\pm0.01$ & $18.49\pm0.01$ & $18.78\pm0.01$ & 	55088.3 & $18.14\pm0.01$ & $18.51\pm0.01$ & $18.48\pm0.01$ & $18.78\pm0.01$  \\
55092.4 & $18.13\pm0.01$ & $18.53\pm0.02$ & $18.52\pm0.03$ & $18.78\pm0.04$ & 	55092.4 & $18.14\pm0.01$ & $18.53\pm0.01$ & $18.48\pm0.03$ & $18.77\pm0.02$  \\
55093.4 & $18.05\pm0.01$ & $18.58\pm0.05$ & $18.51\pm0.08$ & $18.72\pm0.07$ & 	55093.4 & $18.06\pm0.01$ & $18.59\pm0.05$ & $18.50\pm0.03$ & $18.78\pm0.02$  \\

       \hline
     \end{tabular}
   \end{table}
\end{landscape}

 \begin{landscape}
   \begin{table}
     \centering
     \caption{\label{rstars} Light curves in the $R$ filter}
     \begin{tabular}{cccccccccc}
       \hline
       \hline
       \multicolumn{5}{c}{Difference imaging}  &  \multicolumn{5}{c}{PSF fitting}  \\
       MJD & A & B & C & D                       &  MJD & A & B & C & D \\
       \hline
       54675.4 & $17.96\pm0.01$ & $18.45\pm0.02$ & $18.41\pm0.04$ & $18.64\pm0.01$ & 	54675.4 & $17.97\pm0.01$ & $18.47\pm0.02$ & $18.41\pm0.02$ & $18.63\pm0.01$  \\ 
54677.4 & $17.97\pm0.01$ & $18.45\pm0.01$ & $18.42\pm0.01$ & $18.67\pm0.01$ & 	54677.4 & $17.98\pm0.01$ & $18.47\pm0.01$ & $18.43\pm0.01$ & $18.66\pm0.01$  \\ 
54678.4 & $18.03\pm0.01$ & $18.52\pm0.01$ & $18.47\pm0.01$ & $18.73\pm0.01$ & 	54678.4 & $18.03\pm0.01$ & $18.53\pm0.01$ & $18.48\pm0.01$ & $18.73\pm0.01$  \\ 
54681.4 & $17.99\pm0.01$ & $18.47\pm0.01$ & $18.45\pm0.01$ & $18.70\pm0.03$ & 	54681.4 & $17.97\pm0.01$ & $18.48\pm0.01$ & $18.44\pm0.01$ & $18.67\pm0.02$  \\ 
54682.4 & $17.98\pm0.01$ & $18.44\pm0.01$ & $18.41\pm0.03$ & $18.69\pm0.02$ & 	54682.4 & $17.98\pm0.01$ & $18.46\pm0.01$ & $18.42\pm0.01$ & $18.68\pm0.01$  \\ 
54683.4 & $17.95\pm0.01$ & $18.47\pm0.01$ & $18.42\pm0.03$ & $18.66\pm0.01$ & 	54683.4 & $17.97\pm0.01$ & $18.48\pm0.01$ & $18.42\pm0.04$ & $18.64\pm0.01$  \\ 
54684.4 & $17.96\pm0.01$ & $18.45\pm0.01$ & $18.41\pm0.01$ & $18.66\pm0.01$ & 	54684.4 & $17.97\pm0.01$ & $18.46\pm0.01$ & $18.42\pm0.01$ & $18.65\pm0.01$  \\ 
54685.4 & $17.98\pm0.01$ & $18.48\pm0.02$ & $18.43\pm0.01$ & $18.68\pm0.01$ & 	54685.4 & $17.99\pm0.01$ & $18.49\pm0.01$ & $18.43\pm0.01$ & $18.68\pm0.01$  \\ 
54686.4 & $18.00\pm0.01$ & $18.48\pm0.01$ & $18.44\pm0.01$ & $18.70\pm0.02$ & 	54686.4 & $17.99\pm0.01$ & $18.49\pm0.01$ & $18.43\pm0.01$ & $18.70\pm0.01$  \\ 
54688.4 & $17.96\pm0.01$ & $18.46\pm0.01$ & $18.42\pm0.01$ & $18.68\pm0.02$ & 	54688.4 & $17.97\pm0.01$ & $18.47\pm0.02$ & $18.42\pm0.01$ & $18.67\pm0.02$  \\ 
54691.4 & $18.00\pm0.01$ & $18.52\pm0.01$ & $18.43\pm0.05$ & $18.71\pm0.04$ & 	54691.4 & $17.99\pm0.01$ & $18.48\pm0.03$ & $18.40\pm0.01$ & $18.71\pm0.02$  \\ 
54694.4 & $17.99\pm0.01$ & $18.50\pm0.01$ & $18.45\pm0.01$ & $18.68\pm0.01$ & 	54694.4 & $17.99\pm0.01$ & $18.53\pm0.01$ & $18.41\pm0.01$ & $18.69\pm0.01$  \\ 
54698.4 & $17.95\pm0.01$ & $18.45\pm0.01$ & $18.39\pm0.02$ & $18.67\pm0.02$ & 	54698.4 & $17.96\pm0.01$ & $18.46\pm0.01$ & $18.36\pm0.01$ & $18.66\pm0.02$  \\ 
54700.4 & $17.97\pm0.01$ & $18.48\pm0.02$ & $18.40\pm0.02$ & $18.72\pm0.02$ & 	54700.4 & $17.98\pm0.01$ & $18.49\pm0.01$ & $18.39\pm0.01$ & $18.70\pm0.01$  \\ 
54703.3 & $17.99\pm0.01$ & $18.46\pm0.01$ & $18.39\pm0.01$ & $18.66\pm0.01$ & 	54703.3 & $17.99\pm0.01$ & $18.49\pm0.01$ & $18.40\pm0.01$ & $18.68\pm0.01$  \\ 
54704.3 & $17.91\pm0.01$ & $18.40\pm0.01$ & $18.36\pm0.01$ & $18.67\pm0.01$ & 	54704.3 & $17.92\pm0.01$ & $18.44\pm0.01$ & $18.36\pm0.01$ & $18.63\pm0.01$  \\ 
54708.3 & $17.96\pm0.01$ & $18.45\pm0.01$ & $18.38\pm0.01$ & $18.69\pm0.03$ & 	54708.3 & $17.96\pm0.01$ & $18.47\pm0.01$ & $18.37\pm0.01$ & $18.68\pm0.02$  \\ 
54709.4 & $17.92\pm0.01$ & $18.45\pm0.07$ & $18.41\pm0.09$ & $18.63\pm0.07$ & 	54709.4 & $18.01\pm0.01$ & $18.60\pm0.04$ & $18.45\pm0.16$ & $18.75\pm0.01$  \\ 
54720.3 & $17.99\pm0.01$ & $18.46\pm0.01$ & $18.42\pm0.02$ & $18.69\pm0.01$ & 	54720.3 & $17.98\pm0.01$ & $18.45\pm0.01$ & $18.39\pm0.03$ & $18.66\pm0.01$  \\ 
54721.3 & $17.95\pm0.01$ & $18.41\pm0.03$ & $18.37\pm0.03$ & $18.61\pm0.04$ & 	54721.3 & $17.97\pm0.01$ & $18.42\pm0.01$ & $18.40\pm0.02$ & $18.61\pm0.02$  \\ 
54724.3 & $17.99\pm0.01$ & $18.44\pm0.02$ & $18.38\pm0.02$ & $18.64\pm0.01$ & 	54724.3 & $18.01\pm0.01$ & $18.47\pm0.01$ & $18.40\pm0.02$ & $18.65\pm0.01$  \\ 
54725.3 & $17.96\pm0.01$ & $18.43\pm0.04$ & $18.37\pm0.02$ & $18.62\pm0.04$ & 	54725.3 & $17.97\pm0.01$ & $18.47\pm0.02$ & $18.38\pm0.01$ & $18.65\pm0.02$  \\ 
54726.3 & $17.91\pm0.01$ & $18.42\pm0.02$ & $18.34\pm0.02$ & $18.61\pm0.02$ & 	54726.3 & $17.93\pm0.01$ & $18.45\pm0.01$ & $18.35\pm0.02$ & $18.60\pm0.02$  \\ 
54728.3 & $17.96\pm0.01$ & $18.43\pm0.03$ & $18.37\pm0.02$ & $18.63\pm0.02$ & 	54728.3 & $17.99\pm0.01$ & $18.43\pm0.02$ & $18.36\pm0.01$ & $18.62\pm0.01$  \\ 
54729.4 & $17.93\pm0.01$ & $18.44\pm0.01$ & $18.39\pm0.04$ & $18.55\pm0.01$ & 	54729.4 & $17.95\pm0.01$ & $18.44\pm0.01$ & $18.38\pm0.03$ & $18.55\pm0.02$  \\ 
54730.4 & $17.97\pm0.01$ & $18.45\pm0.03$ & $18.38\pm0.02$ & $18.64\pm0.02$ & 	54730.4 & $17.99\pm0.01$ & $18.47\pm0.01$ & $18.39\pm0.01$ & $18.65\pm0.01$  \\ 
54731.4 & $17.98\pm0.01$ & $18.46\pm0.01$ & $18.41\pm0.01$ & $18.66\pm0.02$ & 	54731.4 & $17.98\pm0.01$ & $18.45\pm0.01$ & $18.39\pm0.01$ & $18.65\pm0.01$  \\ 
54733.3 & $18.01\pm0.01$ & $18.47\pm0.02$ & $18.41\pm0.01$ & $18.66\pm0.02$ & 	54733.3 & $18.01\pm0.01$ & $18.47\pm0.01$ & $18.40\pm0.02$ & $18.65\pm0.01$  \\ 
54734.4 & $17.98\pm0.01$ & $18.42\pm0.01$ & $18.38\pm0.01$ & $18.63\pm0.02$ & 	54734.4 & $17.99\pm0.01$ & $18.43\pm0.02$ & $18.38\pm0.03$ & $18.63\pm0.01$  \\ 
54740.4 & $17.97\pm0.01$ & $18.43\pm0.02$ & $18.39\pm0.03$ & $18.65\pm0.02$ & 	54740.4 & $17.99\pm0.01$ & $18.46\pm0.01$ & $18.38\pm0.02$ & $18.64\pm0.01$  \\ 
54741.4 & $17.95\pm0.01$ & $18.41\pm0.01$ & $18.38\pm0.02$ & $18.63\pm0.03$ & 	54741.4 & $17.97\pm0.01$ & $18.43\pm0.01$ & $18.39\pm0.02$ & $18.63\pm0.02$  \\ 
54743.3 & $17.96\pm0.01$ & $18.43\pm0.01$ & $18.40\pm0.01$ & $18.63\pm0.01$ & 	54743.3 & $17.97\pm0.01$ & $18.44\pm0.02$ & $18.39\pm0.01$ & $18.63\pm0.01$  \\ 
55064.4 & $18.32\pm0.01$ & $18.73\pm0.01$ & $18.67\pm0.01$ & $18.92\pm0.02$ & 	55064.4 & $18.30\pm0.01$ & $18.71\pm0.01$ & $18.64\pm0.02$ & $18.92\pm0.01$  \\ 
55065.4 & $18.30\pm0.01$ & $18.71\pm0.01$ & $18.69\pm0.01$ & $18.91\pm0.01$ & 	55065.4 & $18.29\pm0.01$ & $18.69\pm0.01$ & $18.67\pm0.01$ & $18.86\pm0.02$  \\ 
55068.4 & $18.30\pm0.01$ & $18.71\pm0.01$ & $18.67\pm0.01$ & $18.94\pm0.02$ & 	55068.4 & $18.29\pm0.01$ & $18.70\pm0.01$ & $18.65\pm0.01$ & $18.94\pm0.01$  \\ 
55070.4 & $18.31\pm0.01$ & $18.71\pm0.02$ & $18.66\pm0.03$ & $18.90\pm0.01$ & 	55070.4 & $18.30\pm0.01$ & $18.71\pm0.01$ & $18.66\pm0.01$ & $18.90\pm0.01$  \\ 
55071.4 & $18.32\pm0.01$ & $18.72\pm0.01$ & $18.68\pm0.01$ & $18.96\pm0.01$ & 	55071.4 & $18.30\pm0.01$ & $18.70\pm0.01$ & $18.66\pm0.01$ & $18.93\pm0.01$  \\ 
55072.4 & $18.30\pm0.01$ & $18.71\pm0.01$ & $18.64\pm0.02$ & $18.90\pm0.01$ & 	55072.4 & $18.31\pm0.01$ & $18.72\pm0.01$ & $18.64\pm0.02$ & $18.93\pm0.01$  \\ 
55073.4 & $18.33\pm0.01$ & $18.71\pm0.03$ & $18.71\pm0.02$ & $18.94\pm0.02$ & 	55073.4 & $18.31\pm0.01$ & $18.70\pm0.01$ & $18.67\pm0.01$ & $18.94\pm0.01$  \\ 
55074.4 & $18.31\pm0.01$ & $18.69\pm0.01$ & $18.66\pm0.02$ & $18.92\pm0.01$ & 	55074.4 & $18.31\pm0.01$ & $18.69\pm0.01$ & $18.67\pm0.01$ & $18.94\pm0.03$  \\ 
55079.4 & $18.38\pm0.01$ & $18.72\pm0.08$ & $18.71\pm0.03$ & $19.00\pm0.05$ & 	55079.4 & $18.35\pm0.01$ & $18.70\pm0.04$ & $18.69\pm0.07$ & $18.98\pm0.04$  \\ 
55085.4 & $18.31\pm0.01$ & $18.70\pm0.04$ & $18.66\pm0.01$ & $18.89\pm0.02$ & 	55085.4 & $18.29\pm0.01$ & $18.70\pm0.01$ & $18.64\pm0.03$ & $18.92\pm0.01$  \\ 
55087.4 & $18.33\pm0.01$ & $18.73\pm0.03$ & $18.66\pm0.03$ & $18.96\pm0.03$ & 	55087.4 & $18.32\pm0.01$ & $18.69\pm0.01$ & $18.64\pm0.04$ & $18.90\pm0.02$  \\ 
55088.3 & $18.32\pm0.01$ & $18.69\pm0.01$ & $18.66\pm0.01$ & $18.93\pm0.01$ & 	55088.3 & $18.31\pm0.01$ & $18.68\pm0.01$ & $18.65\pm0.02$ & $18.91\pm0.02$  \\ 
55092.4 & $18.33\pm0.01$ & $18.75\pm0.02$ & $18.69\pm0.01$ & $18.95\pm0.02$ & 	55092.4 & $18.33\pm0.01$ & $18.72\pm0.02$ & $18.65\pm0.01$ & $18.92\pm0.01$  \\ 
55093.4 & $18.32\pm0.01$ & $18.70\pm0.02$ & $18.67\pm0.01$ & $18.94\pm0.04$ & 	55093.4 & $18.31\pm0.01$ & $18.70\pm0.01$ & $18.66\pm0.01$ & $18.95\pm0.01$  \\ 

       \hline
     \end{tabular}
   \end{table}
\end{landscape}

 \begin{landscape}
   \begin{table}
     \centering
     \caption{\label{vstars} Light curves in the $V$ filter}
     \begin{tabular}{cccccccccc}
       \hline
       \hline
       \multicolumn{5}{c}{Difference imaging}  &  \multicolumn{5}{c}{PSF fitting}  \\
       MJD & A & B & C & D                       &  MJD & A & B & C & D \\
       \hline
       54675.4 & $18.45\pm0.01$ & $18.96\pm0.01$ & $18.94\pm0.04$ & $19.12\pm0.03$ & 	54675.4 & $18.45\pm0.01$ & $18.93\pm0.02$ & $18.92\pm0.01$ & $19.14\pm0.03$ \\
54677.4 & $18.45\pm0.01$ & $18.97\pm0.01$ & $18.94\pm0.01$ & $19.13\pm0.01$ & 	54677.4 & $18.44\pm0.01$ & $18.95\pm0.01$ & $18.91\pm0.01$ & $19.14\pm0.02$ \\
54681.4 & $18.45\pm0.01$ & $18.99\pm0.01$ & $18.92\pm0.01$ & $19.16\pm0.01$ & 	54681.4 & $18.49\pm0.01$ & $18.98\pm0.02$ & $18.95\pm0.02$ & $19.18\pm0.01$ \\
54682.4 & $18.48\pm0.01$ & $18.98\pm0.01$ & $18.93\pm0.02$ & $19.13\pm0.02$ & 	54682.4 & $18.48\pm0.01$ & $18.98\pm0.01$ & $18.92\pm0.01$ & $19.14\pm0.01$ \\
54683.4 & $18.47\pm0.01$ & $19.01\pm0.03$ & $18.92\pm0.04$ & $19.13\pm0.03$ & 	54683.4 & $18.46\pm0.01$ & $18.99\pm0.03$ & $18.90\pm0.01$ & $19.14\pm0.02$ \\
54684.4 & $18.47\pm0.01$ & $18.98\pm0.02$ & $18.94\pm0.02$ & $19.15\pm0.01$ & 	54684.4 & $18.47\pm0.01$ & $18.97\pm0.02$ & $18.92\pm0.02$ & $19.15\pm0.02$ \\
54685.4 & $18.48\pm0.01$ & $18.99\pm0.01$ & $18.93\pm0.01$ & $19.15\pm0.02$ & 	54685.4 & $18.48\pm0.01$ & $18.99\pm0.01$ & $18.93\pm0.02$ & $19.15\pm0.04$ \\
54686.4 & $18.46\pm0.01$ & $19.00\pm0.01$ & $18.93\pm0.01$ & $19.16\pm0.01$ & 	54686.4 & $18.47\pm0.01$ & $18.99\pm0.02$ & $18.92\pm0.01$ & $19.15\pm0.01$ \\
54690.4 & $18.46\pm0.01$ & $18.99\pm0.01$ & $18.92\pm0.02$ & $19.16\pm0.02$ & 	54690.4 & $18.46\pm0.01$ & $18.97\pm0.01$ & $18.88\pm0.01$ & $19.16\pm0.02$ \\
54698.4 & $18.45\pm0.01$ & $18.96\pm0.01$ & $18.90\pm0.01$ & $19.15\pm0.01$ & 	54698.4 & $18.48\pm0.01$ & $18.99\pm0.02$ & $18.93\pm0.02$ & $19.18\pm0.01$ \\
54700.4 & $18.47\pm0.01$ & $18.99\pm0.02$ & $18.89\pm0.01$ & $19.16\pm0.01$ & 	54700.4 & $18.47\pm0.01$ & $18.98\pm0.01$ & $18.89\pm0.01$ & $19.17\pm0.01$ \\
54703.4 & $18.45\pm0.01$ & $18.99\pm0.01$ & $18.91\pm0.03$ & $19.14\pm0.01$ & 	54703.4 & $18.44\pm0.01$ & $18.96\pm0.01$ & $18.89\pm0.04$ & $19.14\pm0.01$ \\
54708.3 & $18.46\pm0.01$ & $18.97\pm0.01$ & $18.88\pm0.02$ & $19.16\pm0.01$ & 	54708.3 & $18.45\pm0.01$ & $18.94\pm0.01$ & $18.87\pm0.01$ & $19.16\pm0.02$ \\
54720.3 & $18.46\pm0.01$ & $18.95\pm0.01$ & $18.87\pm0.01$ & $19.09\pm0.03$ & 	54720.3 & $18.46\pm0.01$ & $18.92\pm0.03$ & $18.85\pm0.01$ & $19.08\pm0.01$ \\
54724.3 & $18.49\pm0.01$ & $18.95\pm0.01$ & $18.89\pm0.03$ & $19.14\pm0.01$ & 	54724.3 & $18.47\pm0.01$ & $18.94\pm0.05$ & $18.86\pm0.01$ & $19.11\pm0.03$ \\
54725.3 & $18.48\pm0.01$ & $18.95\pm0.01$ & $18.91\pm0.01$ & $19.16\pm0.03$ & 	54725.3 & $18.51\pm0.01$ & $18.98\pm0.01$ & $18.93\pm0.01$ & $19.18\pm0.05$ \\
54726.3 & $18.43\pm0.01$ & $18.92\pm0.02$ & $18.86\pm0.01$ & $19.10\pm0.02$ & 	54726.3 & $18.44\pm0.01$ & $18.92\pm0.01$ & $18.87\pm0.01$ & $19.13\pm0.01$ \\
54728.4 & $18.44\pm0.01$ & $18.98\pm0.01$ & $18.89\pm0.01$ & $19.16\pm0.01$ & 	54728.4 & $18.40\pm0.01$ & $18.87\pm0.01$ & $18.78\pm0.01$ & $19.08\pm0.01$ \\
54730.4 & $18.45\pm0.01$ & $18.93\pm0.02$ & $18.89\pm0.01$ & $19.13\pm0.01$ & 	54730.4 & $18.44\pm0.01$ & $18.93\pm0.02$ & $18.88\pm0.04$ & $19.11\pm0.04$ \\
54731.4 & $18.45\pm0.01$ & $18.94\pm0.01$ & $18.88\pm0.03$ & $19.12\pm0.03$ & 	54731.4 & $18.44\pm0.01$ & $18.93\pm0.01$ & $18.85\pm0.02$ & $19.10\pm0.02$ \\
54733.4 & $18.45\pm0.01$ & $18.93\pm0.01$ & $18.87\pm0.02$ & $19.11\pm0.02$ & 	54733.4 & $18.43\pm0.01$ & $18.91\pm0.01$ & $18.85\pm0.01$ & $19.09\pm0.03$ \\
54734.4 & $18.51\pm0.01$ & $18.89\pm0.01$ & $18.88\pm0.01$ & $19.10\pm0.01$ & 	54734.4 & $18.51\pm0.01$ & $18.91\pm0.01$ & $18.87\pm0.01$ & $19.11\pm0.01$ \\
54740.4 & $18.47\pm0.01$ & $18.94\pm0.01$ & $18.89\pm0.03$ & $19.09\pm0.01$ & 	54740.4 & $18.46\pm0.01$ & $18.93\pm0.01$ & $18.88\pm0.02$ & $19.10\pm0.02$ \\
54741.4 & $18.45\pm0.01$ & $18.94\pm0.01$ & $18.88\pm0.02$ & $19.11\pm0.01$ & 	54741.4 & $18.44\pm0.01$ & $18.92\pm0.01$ & $18.87\pm0.01$ & $19.09\pm0.02$ \\
54743.3 & $18.45\pm0.01$ & $18.93\pm0.01$ & $18.86\pm0.01$ & $19.09\pm0.01$ & 	54743.3 & $18.44\pm0.01$ & $18.91\pm0.02$ & $18.84\pm0.01$ & $19.08\pm0.01$ \\
55068.3 & $18.86\pm0.01$ & $19.23\pm0.01$ & $19.19\pm0.01$ & $19.45\pm0.01$ & 	55068.3 & $18.88\pm0.01$ & $19.22\pm0.01$ & $19.17\pm0.02$ & $19.46\pm0.01$ \\
55070.4 & $18.85\pm0.01$ & $19.22\pm0.01$ & $19.20\pm0.01$ & $19.46\pm0.01$ & 	55070.4 & $18.89\pm0.01$ & $19.24\pm0.01$ & $19.20\pm0.01$ & $19.49\pm0.02$ \\
55071.4 & $18.84\pm0.01$ & $19.23\pm0.01$ & $19.20\pm0.02$ & $19.45\pm0.01$ & 	55071.4 & $18.86\pm0.01$ & $19.21\pm0.01$ & $19.19\pm0.03$ & $19.47\pm0.01$ \\
55072.4 & $18.85\pm0.01$ & $19.23\pm0.01$ & $19.20\pm0.01$ & $19.44\pm0.01$ & 	55072.4 & $18.86\pm0.01$ & $19.20\pm0.01$ & $19.18\pm0.03$ & $19.43\pm0.03$ \\
55073.4 & $18.84\pm0.01$ & $19.23\pm0.01$ & $19.19\pm0.01$ & $19.45\pm0.02$ & 	55073.4 & $18.88\pm0.01$ & $19.23\pm0.01$ & $19.20\pm0.03$ & $19.47\pm0.03$ \\
55074.4 & $18.85\pm0.01$ & $19.24\pm0.03$ & $19.21\pm0.01$ & $19.44\pm0.01$ & 	55074.4 & $18.88\pm0.01$ & $19.22\pm0.03$ & $19.20\pm0.01$ & $19.43\pm0.01$ \\
55075.4 & $18.85\pm0.01$ & $19.23\pm0.01$ & $19.18\pm0.03$ & $19.46\pm0.01$ & 	55075.4 & $18.88\pm0.01$ & $19.23\pm0.02$ & $19.18\pm0.04$ & $19.46\pm0.01$ \\
55085.4 & $18.86\pm0.01$ & $19.24\pm0.01$ & $19.20\pm0.01$ & $19.44\pm0.01$ & 	55085.4 & $18.88\pm0.01$ & $19.22\pm0.04$ & $19.19\pm0.04$ & $19.45\pm0.03$ \\
55087.4 & $18.85\pm0.01$ & $19.26\pm0.01$ & $19.21\pm0.01$ & $19.39\pm0.01$ & 	55087.4 & $18.86\pm0.01$ & $19.23\pm0.01$ & $19.20\pm0.01$ & $19.43\pm0.01$ \\
55088.3 & $18.86\pm0.01$ & $19.22\pm0.01$ & $19.20\pm0.02$ & $19.44\pm0.02$ & 	55088.3 & $18.90\pm0.01$ & $19.22\pm0.01$ & $19.19\pm0.02$ & $19.47\pm0.02$ \\
55092.3 & $18.85\pm0.01$ & $19.24\pm0.01$ & $19.20\pm0.01$ & $19.46\pm0.02$ & 	55092.3 & $18.86\pm0.01$ & $19.23\pm0.03$ & $19.21\pm0.04$ & $19.46\pm0.03$ \\
55093.4 & $18.85\pm0.01$ & $19.24\pm0.01$ & $19.19\pm0.02$ & $19.45\pm0.01$ & 	55093.4 & $18.88\pm0.01$ & $19.24\pm0.01$ & $19.20\pm0.02$ & $19.46\pm0.01$ \\

       \hline
     \end{tabular}
   \end{table}
\end{landscape}

\end{document}